# Using the knowledge of penumbra with a trick simulation

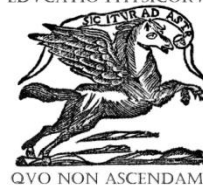

**Mustafa Şahin Bülbül[1], Loo Kang Wee[2]**
[1]*Faculty of Education, Kafkas University, Turkey.*
[2]*Ministry of Education, Education Technology Division, Singapore.*

**E-mail**: msahinbulbul@gmail.com

**Abstract**
The study is about a basic shadow experiment, which was enriched with a simulation to understand the reasoning of participants when we use a trick. Two light sources create an umbra and penumbra behind the objects. With this experiment, we asked what would happen when the penumbras interact. Most of the participants predicted the correct solution, that there should be an umbra. Some of the participants choose wrong alternative, and explained in terms of the structure of penumbra.

**Resumen**
El estudio trata de un experimento básico con una sombra que fue enriquecida con una simulación, para entender el razonamiento de los participantes cuando utilizamos un truco. Dos fuentes de luz crearon umbra y penumbra detrás de los objetos. Con este experimento nos preguntamos qué pasará cuando las penumbras interactúan. La mayoría de los participantes predijeron la solución correcta, que debe haber una umbra. Algunos de los participantes eligen la alternativa equivocada, y se explica en términos de la estructura de la penumbra.

**Palabras clave:** Simulación fácil de Java, Penumbras, Aprendizaje activo, Desarrollo profesional de los docentes, E-learning, Applets, Diseño, Código abierto, Física de Nivel Avanzado GCE.

## I. INTRODUCTION

Different experiences and problems about shadow generally take students' interest and make the learning process easier.

For instance, Hughes [1] used software and analyzed the image of white paper under a shadow. To measure the radius of the Earth by using the shadow of building [2], or to observe the shadow of virtual images [3], are some of other interesting exercises about shadow. This study is prepared to bring forward the intersection between two penumbras.

This study focuses on the penumbra regions, B1 and B2 concept in the phenomena of light and shadow using a simple real life setup experiment. There are two light sources labeled S1 and S2 (0) and two opaque rectangular wooden block objects to create shadows.

The penumbra concept is ambiguous for students, as it could be thought partly lighted (semi-lighted) or partly shadowed (semi-shadowed) regions, depending on their light or shadow concept perspectives.

Thus, an investigated phenomenon in this study was conducted with pre-service physics teachers, on their predictions about the intersection between two penumbras, or called antumbra, labeled as C.

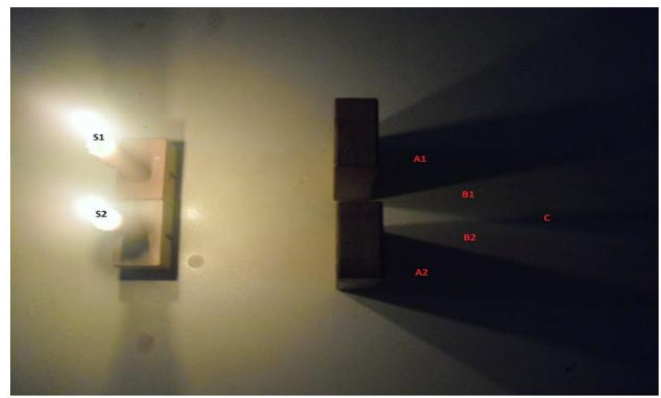

**FIGURE 1.** A simple real life setup using 2 lamps as sources S1 and S2, 2 blocks and the resultant shadow regions of A1 umbra - darkest part of the shadow, B1, B2 penumbra - semi-lighted or semi-shadowedand C antumbra. Picture by M. Ş. Bulbul.

## II. RESEARCH DESIGN

Interviews with N=15 pre-service teacher trainees were conducted where three different question papers were prepared for this investigation. Each paper included two questions: one for the student prediction, and one for the reason of their prediction.

After these two questions, it was asked whether they are sure or not. This question used to understand how they are sure of their answer.

The question to understand the pre-service teachers' prediction was designed in three types; focused on sources, slit or the intersection between penumbras of two different light sources. For this question there were three alternatives as penumbra, lighted or shadowed (umbra).

## III. LIGHT AND SHADOW COMPUTER MODEL

The main point of this simulation is to show that light rays leaving an object travel in straight lines. The simulation has two light bulbs that can be turned on or off independently.

The bulbs can be dragged around the screen to change their positions (Figure 2). Either the light from the bulbs passes through a mask with a hole in it, or it is blocked by an object, as it travels to a screen on the right of the simulation.

Users may investigate how the patterns of light and shadows change as they move the bulbs, the mask or object, and/or the screen, and as they change the size of the mask or object.

By selecting the "what's wrong" section, the intersection area of two penumbras turns into illuminated. In natural setting, the interaction area of two penumbras should be shadowed; however, selecting the "what's wrong" section makes a trick. This trick simulation used to change the pre-service teachers' view that had chosen umbra and explained that s/he is not sure.

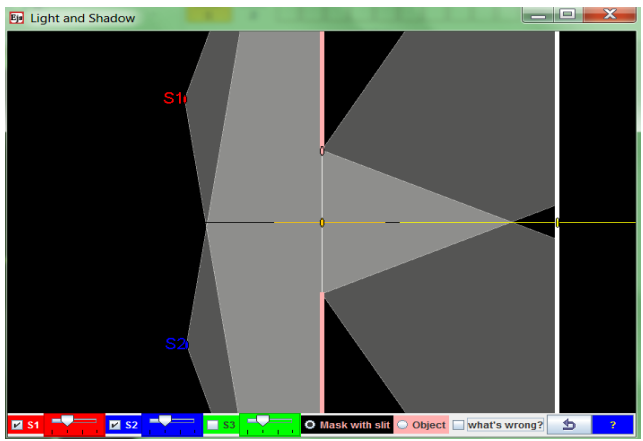

**FIGURE 2.** Light and shadow computer model view of the similar setup with light sources S1 and S2, draggable mask-opaque blocks (pink) with slit with the correct representation of the intersection of B1 and B2 as antumbra C.

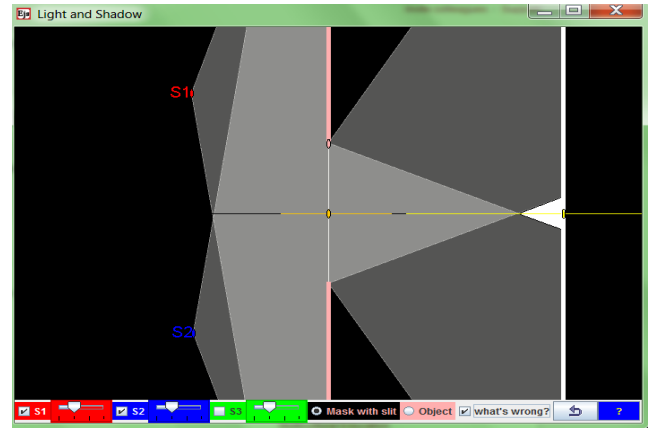

**FIGURE 3.** Light and shadow computer model view of the similar setup with light sources S1 and S2, draggable mask-opaque blocks (pink) with slit with the incorrect representation of the intersection of B1 and B2 as lighted region.

## IV. RESULTS

Ten students, which their questions is focused on slit or light sources, they said there will be shadow with different explanations; however, three students in the third group said that there will not be a shadow. For instance, one participant explained that there will be an unreached area by the light of two sources, so there will be a shadow and in the second group where the question is designed on slits, one of the participant explained that less light entrance is the reason why closer slits make shadow.

On the other hand, in the third group, one of the participants explained that interaction area includes more light from different penumbras. The other participants explain the reason why they do not predict shadow that there is no barrier in front of the interaction area.

## V. AFTER THE INTERVIEW

Among the subjects who answered the question focused on bulbs or slit were successful to explain the event. However, all the subjects who answered the question focused on the intersection of penumbras additionally not sure pre-service teachers selected the correct alternative explained that penumbras include particle of lights and that area should include these particles more.

The question was easy for pre-service teachers who answered "shadow" because light sources diverge from each other or smaller slit are two factors affect that area. Although the phenomenon is same, the focused of question and the trick simulation changed pre-service teachers' perspective.

They explained that if the area illuminated by two bulbs is brighter than the area illuminated by one bulb, the area of intersection between two penumbras should be brighter than the area of any penumbra.

## VI. CONCLUSION

Shadow is a situation of light and object, so penumbra phenomenon should be evaluated in terms of these primary factors. If an instructor, ask the question through the secondary factors as this study subjects may prefer to think the penumbra as it is a source of light.

Similarly, if the simulation reflects the wrong phenomenon, subjects may use the same reasoning. The simulation is available as a downloadable jar file from: http://weelookang.blogspot.com/2013/08/light-and-shadow-model.html, and from: http://weelookang.blogspot.com/2013/09/light-and-shadow-model-with-trick.html.

## ACKNOWLEDGEMENT

We wish to thank Andrew Duffy for his original computer model, that serves as a template from which we derived our customized Light and Shadow Model. We wish to acknowledge the passionate contributions of Francisco Esquembre, Fu-Kwun Hwang, and Wolfgang Christian for their ideas and insights in the co-creation of interactive simulation and curriculum materials.